# Realization of *t*-bit semiclassical quantum Fourier transform on IBM's quantum cloud computer

Fu Xiang-qun[1,2], Bao Wan-su[1,2,*], Huang He-liang[1,2], Li Tan[1,2], Shi Jian-hong[1,2], Wang Xiang[1,2], Zhang Shuo[1,2] and Li Feng-guang[1,2]

[1]*Henan Key Laboratory of Quantum Information and Cryptography, Zhengzhou, Henan 450004, China*

[2]*Synergetic Innovation Center of Quantum Information and Quantum Physics, University of Science and Technology of China, Hefei, Anhui 230026, China*

**Correspondence to: 2010thzz@sina.com*

**Abstract:** To overcome the difficulty of realizing large-scale quantum Fourier transform (QFT) within existing technology, this paper presents a resource-saving method, namely *t*-bit semiclassical QFT over $Z_{2^n}$, which could realize large-scale QFT using arbitrary-scale quantum register. Using our method, the scale of quantum register can be determined flexibility according to the scale of quantum system, enabling the quantum resource and speed of realizing QFT to be optimal. By developing a feasible method to realize the control quantum gate $R_k$, we experimentally demonstrate the 2-bit semiclassical QFT over $Z_{2^3}$ on IBM's quantum cloud computer, showing the feasibility of our proposed method. Then, we compare the actual performance of 2-bit semiclassical QFT and standard QFT in the experiments. Experimental data show that the fidelity of the result of 2-bit semiclassical QFT is higher than that of standard QFT, which is mainly due to less two-qubit controlled gates are required in the semiclassical QFT. Furthermore, based on the proposed method, we successfully implement the Shor's algorithm to factorize *N*=15.

## 1 Introduction

The quantum algorithm[1][2] greatly threatens the current cryptography, especially the public-key cryptography, whose security is based on the factorization and discrete logarithm[3][4], which can be easily broken by Shor's algorithm. And great attention has since been paid to research on quantum computation [5]-[10]. Usually, the quantum algorithm is composed of different quantum transforms, where the quantum Fourier transform (QFT) is one of the most important transforms. It can be used not only for preparing the equal superposition state, but also for increasing the success probability of the quantum algorithm. In 1994, D.Coppersmith [11] presented the circuit for the QFT over $Z_{2^n}$, which needs a $n$-qubit quantum register, $n$ single-qubit gates and $n(n+2)/2$ two-qubit gates. Thus, it is very resource-consuming for realizing the QFT over $Z_{2^n}$ for larger $n$ using this scheme.

In 1996, R.B.Griffiths and C.Niu[12] presented the single-bit semiclassical QFT over $Z_{2^n}$, which needs only an single-qubit quantum register and 2 single-qubit gates. Thus, in their scheme, not only the quantum computational resource but also the implementation difficulty is reduced (two-qubit controlled gate is more difficult to realize than single-qubit gate in experiments). And Chiaverini et al.[13] experimentally implemented this kind of the single-bit semiclassical QFT over $Z_{2^3}$ on the ion-trap system in 2005, which demonstrated the feasibility of the semiclassical QFT. In 2000, based on the single-bit semiclassical QFT, Parker et al.[14] presented a circuit for the $n+1$-qubit quantum register for factorization problem. While, based on the QFT, $3n+O(\log N)$-qubit quantum register is needed[15]. Thus, the single-bit semiclassical QFT provides a solution for implementing the large-scale QFT, with which Shor's algorithm can be applied to break the public-key cryptology (e.g., RSA and ECC) under the condition of limited computation resources.

Considering the two-qubit gate is more difficult to implement than the one-qubit gate, thus the two-qubit gate should be avoided as many as possible in a quantum computation circuit. The one-bit semiclassical QFT over $Z_{2^n}$, compared with the QFT over $Z_{2^n}$, needs less quantum computation resource, thus to replace the QFT with a one-bit semiclassical QFT is a good choice. However, it also sacrifices the implementation speed. Due to the difficulty of realizing a large-scale quantum register, how to balance the trade-off between quantum resource and speed in the implementation of the QFT (e.g., how to implement a large-scale QFT based on a small-scale quantum register) requires further studies.

In this paper, we define the *t*-bit semiclassical QFT over $Z_{2^n}$, which can realize QFT. And the requirements for the two-qubit and single-qubit gates are respectively $1/l^2$ and $2/l$ times those of the QFT over $Z_{2^n}$ (if $t$ divides $n$ exactly, $l=[n/t]-1$, otherwise $l=[n/t]$), and clarify that the speed of the *t*-bit semiclassical QFT increases to $t$ times that of the single-bit case. Finally, to explore the actual performance of QFT, we design and realize the 2-bit semiclassical QFT over $Z_{2^3}$ on IBM's quantum cloud quantum computer. And the method may provide a feasible route to realize the QFT over $Z_{2^n}$.

## 2 The QFT over $Z_{2^n}$

**Definition1** (quantum Fourier transform)[15] The QFT on an orthonormal basis $|0\rangle,|1\rangle,\cdots,|N-1\rangle$ is defined to be a linear operator $U_F$ with the following action on the basis states,

$$U_F:|j\rangle \mapsto \frac{1}{\sqrt{N}}\sum_{k=0}^{N-1}\omega_N^{jk}|k\rangle .$$

So the action on an arbitrary state can be written as

$$U_F:\sum_{j=0}^{N-1}x_j|j\rangle \mapsto \frac{1}{\sqrt{N}}\sum_{k=0}^{N-1}\sum_{j=0}^{N-1}x_j\omega_N^{jk}|k\rangle .$$

where $i=\sqrt{-1}$, $\omega_N=e^{2\pi i/N}$ and the symbol "$\mapsto$" stands for that the transform is invertible.

When $N=2^n$, the QFT $U_F$ on the state $|a\rangle$ is given as

$$U_F|a\rangle=\bigotimes_{j=0}^{n-1}\left|p(\phi_j)\right\rangle_j$$

where $U_F$ is also called the QFT over $Z_{2^n}$, $a=\sum_{j=0}^{n-1}a_j2^j$ in the binary representation ($a_j$ is 0 or 1), $|p(\phi)\rangle=(|0\rangle+e^{2\pi i\phi}|1\rangle)/\sqrt{2}$ and $\phi_j=\sum_{k=0}^{n-j}a_k2^{j+k-n-1}$ for $j=0,1,\cdots,n-1$.

The QFT is a usual transform in quantum algorithms, which can be applied to prepare the superposition state. The circuit for the QFT over $Z_{2^n}$ needs a $n$-qubit quantum register, $n$ single-qubit gates and $n(n+2)/2$ two-qubit gates, which is designed by D.Coppersmith [11]. And the circuit is called standard QFT.

Based on the standard QFT, the QFT over $Z_{2^n}$ can't be realized within existing technology for larger $n$. In 1996, R.B.Griffiths and C.Niu[12] presented a method that only required a single-qubit quantum register and 2 single-qubit gates to implement the QFT, which was called single-bit semiclassical QFT over $Z_{2^n}$.

Single-bit semiclassical QFT over $Z_{2^n}$ carries each basis state $|a\rangle=|a_{n-1},\cdots,a_0\rangle$ into $|c\rangle=|c_{n-1},\cdots,c_0\rangle$ whose amplitude is equal to that of QFT[12]. Thus, single-bit semiclassical QFT over $Z_{2^n}$ can realize QFT and $|a\rangle$'s input is from left to right by bit flipping. That is to say, single-bit semiclassical QFT over $Z_{2^n}$ can carry out the parallel computing of $2^n$ inputs within $n$ steps. Compared to standard QFT, through measurement after transforming each bit, single-bit semiclassical QFT over $Z_{2^n}$ requires less quantum computation resource. And the quantum register with 1 qubit can be applied cyclically to prepare $|a_{n-1}\rangle,\cdots,|a_0\rangle$ [14].

## 3 *t*-bit semiclassical QFT over $Z_{2^n}$

The standard QFT has a higher implementation speed but needs more quantum computation resources, whereas the single-bit semiclassical QFT needs fewer resources but has lower speed. To overcome the disadvantages of the two methods, we design a new realization method for QFT to balance the tradeoffs of the quantum resource and speed.

**Definition 2** Suppose $t$ is a positive integer and $l=\begin{cases}[n/t]\ ,\ t\nmid n\\ l=[n/t]-1\ ,\ t|n\end{cases}$, $|a\rangle=|a_{n-1},a_{n-2},\cdots,a_0\rangle$ is an arbitrary

orthonormal state with orthonormal basis $|0\rangle,|1\rangle,\cdots,|N-1\rangle$. If $U_{F_t}$ acts on a block of bits of $|a\rangle$ as

$$|a_{n-1},a_{n-2},\cdots,a_{lt}\rangle \mapsto \frac{1}{2^{(n-lt)/2}}\sum_{c_0'=0}^{2^{n-lt}-1}\omega_{2^{n-lt}}^{a_0'c_0'}|c_{n-lt-1},c_{n-lt-2},\cdots,c_0\rangle,$$

$$|a_{lt-1},a_{lt-2},\cdots,a_{(l-1)t}\rangle \mapsto \frac{1}{2^{t/2}}\sum_{c_1'=0}^{2^t-1}\omega_{2^t}^{a_1'c_1'}e^{2\pi i(a_{lt-1}+\frac{a_{lt-2}}{2}+\cdots+\frac{a_{(l-1)t}}{2^{t-1}})\varphi_1}|c_{n-(l-1)t-1},c_{n-(l-1)t-2},\cdots,c_{n-lt}\rangle$$

$$\vdots$$

$$|a_{t-1},a_{t-2},\cdots,a_0\rangle \mapsto \frac{1}{2^{t/2}}\sum_{c_l'=0}^{2^t-1}\omega_{2^t}^{a_l'c_l'}e^{2\pi i(a_{t-1}+\frac{a_{t-2}}{2}+\cdots+\frac{a_0}{2^{t-1}})\varphi_l}|c_{n-1},c_{n-2},\cdots,c_{n-t}\rangle,$$

where $a_0'=\sum_{r=lt}^{n-1}2^{r-lt}a_r$ , $a_j'=\sum_{r=(l-j)t}^{(l-j+1)t-1}2^{r-(l-j)t}a_r$ , $c_0'=\sum_{r=0}^{n-lt-1}2^r c_r$ , $c_j'=\sum_{r=n-(l-j+1)t}^{n-(l-j)t-1}2^{r-n+(l-j+1)t}c_r$ , $\varphi_1=\frac{c_0}{2^{n-lt+1}}+\cdots+\frac{c_{n-lt-1}}{2^2}$ and $\varphi_k=\frac{\varphi_{k-1}}{2^t}+\frac{c_{n-(l-k+2)t}}{2^{t+1}}+\cdots+\frac{c_{n-(l-k+1)t-1}}{2^2}$ ( $j=1,\cdots,l$ , $k=2,\cdots,l$ ), then $U_{F_t}$ is called the *t*-bit semiclassical QFT over $Z_{2^n}$.

In essence, $U_{F_t}$ is a block transform. That is to say, $|a\rangle$ is transformed from left to right in blocks, the first of which are $n-lt$ bits and the other are $t$ bits.

**Theorem 1** Suppose $|0\rangle,|1\rangle,\cdots,|N-1\rangle$ is an orthonormal basis, $U_F$ is the QFT over $Z_{2^n}$ and $U_{F_t}$ is the *t*-bit semiclassical QFT over $Z_{2^n}$. $U_F$ carries $|a\rangle=|a_{n-1},a_{n-2},\cdots,a_0\rangle$ into $|c\rangle=|c_{n-1},c_{n-2},\cdots,c_0\rangle$, whose probability is equal to that of $U_{F_t}$.

**Proof**: To prove the theorem, we only need to prove that $U_F$ carries $|a\rangle=|a_{n-1},a_{n-2},\cdots,a_0\rangle$ into $|c\rangle=|c_{n-1},c_{n-2},\cdots,c_0\rangle$, whose amplitude is equal to that for $U_{F_t}$.

From Definition 2 and $e^{2\pi i}=1$, we obtain

$$\omega_{2^{n-lt}}^{a_0'c_0'}=\omega_{2^{n-lt}}^{a_0'c}=e^{\pi i(\frac{ca_{n-1}}{2^0}+\frac{ca_{n-2}}{2^1}+\cdots+\frac{ca_{lt}}{2^{n-lt-1}})}.$$

Therefore, the amplitude of $|c_{n-lt-1},c_{n-lt-2},\cdots,c_0\rangle$ is $\frac{1}{2^{(n-lt)/2}}e^{\pi i(\frac{ca_{n-1}}{2^0}+\frac{ca_{n-2}}{2^1}+\cdots+\frac{ca_{lt}}{2^{n-lt-1}})}$.

Since

$$\begin{aligned}
&\omega_{2^t}^{a_1'c_1'}e^{2\pi i(a_{lt-1}+\frac{a_{lt-2}}{2}+\cdots+\frac{a_{(l-1)t}}{2^{t-1}})\varphi_1}\\
&=e^{\frac{2\pi i a_1'c_1'}{2^t}}e^{2\pi i(a_{lt-1}+\frac{a_{lt-2}}{2}+\cdots+\frac{a_{(l-1)t}}{2^{t-1}})(\frac{c_0}{2^{n-lt+1}}+\cdots+\frac{c_{n-lt-1}}{2^2})}\\
&=e^{\frac{\pi i a_{lt-1}}{2^{n-lt}}(c_0+2c_1+\cdots+2^{n-(l-1)t-1}c_{n-(l-1)t-1})}\cdots e^{\frac{\pi i a_{(l-1)t}}{2^{n-(l-1)t-1}}(c_0+2c_1+\cdots+2^{n-(l-1)t-1}c_{n-(l-1)t-1})}\\
&=e^{\pi i(\frac{ca_{lt-1}}{2^{n-lt}}+\frac{ca_{lt-2}}{2^{n-lt+1}}+\cdots+\frac{ca_{(l-1)t}}{2^{n-(l-1)t-1}})},
\end{aligned}$$

the amplitude of $|c_{n-(l-1)t-1},c_{n-(l-1)t-2},\cdots,c_{n-lt}\rangle$ is $\frac{1}{2^{t/2}}e^{\pi i(\frac{ca_{lt-1}}{2^{n-lt}}+\frac{ca_{lt-2}}{2^{n-lt+1}}+\cdots+\frac{ca_{(l-1)t}}{2^{n-(l-1)t-1}})}$.

Similarly,

$$\text{the amplitude of } |c_{n-(l-2)t-1},c_{n-(l-2)t-2},\cdots,c_{n-(l-1)t}\rangle \text{ is } \frac{1}{2^{t/2}}e^{\pi i(\frac{ca_{(l-1)t-1}}{2^{n-(l-1)t}}+\frac{ca_{(l-1)t-2}}{2^{n-(l-1)t+1}}+\cdots+\frac{ca_{(l-2)t}}{2^{n-(l-2)t-1}})},$$

$$\vdots$$

$$\text{the amplitude of } |c_{n-1},c_{n-2},\cdots,c_{n-t}\rangle \text{ is } \frac{1}{2^{t/2}}e^{\pi i(\frac{ca_{t-1}}{2^{n-t}}+\frac{ca_{t-2}}{2^{n-t+1}}+\cdots+\frac{ca_0}{2^{n-1}})}.$$

In conclusion, after $U_{F_t}$ acting on $|a\rangle$, the amplitude of $|c\rangle=|c_{n-1},c_{n-2},\cdots,c_0\rangle$ is

$$A=\frac{1}{2^{n/2}}e^{\pi i(\frac{ca_{n-1}}{2^0}+\frac{ca_{n-2}}{2^1}+\cdots+\frac{ca_{lt}}{2^{n-lt-1}})}e^{\pi i(\frac{ca_{lt-1}}{2^{n-lt}}+\frac{ca_{lt-2}}{2^{n-lt+1}}+\cdots+\frac{ca_{(l-1)t}}{2^{n-(l-1)t-1}})}\cdots e^{\pi i(\frac{ca_{t-1}}{2^{n-t}}+\frac{ca_{t-2}}{2^{n-t+1}}+\cdots+\frac{ca_0}{2^{n-1}})}.$$

Furthermore, $A=\frac{1}{2^{n/2}}\omega_{2^n}^{ac}$.

From Definition 1, we obtain

$$U_F\left|a\right\rangle=\frac{1}{2^{n/2}}\sum_{c=0}^{2^n-1}\omega_{2^n}^{ac}\left|c\right\rangle.$$

Then, after $U_F$ acting on $\left|a\right\rangle$, the amplitude of $\left|c\right\rangle=\left|c_{n-1},c_{n-2},\cdots,c_0\right\rangle$ is

$$\frac{1}{2^{n/2}}\omega_{2^n}^{ac}.$$

We thus obtain Theorem 1.

According to Theorem 1, the *t*-bit semiclassical QFT over $Z_{2^n}$ can realize the quantum Fourier transform over $Z_{2^n}$. The circuit for the transform is illustrated in Fig.1.

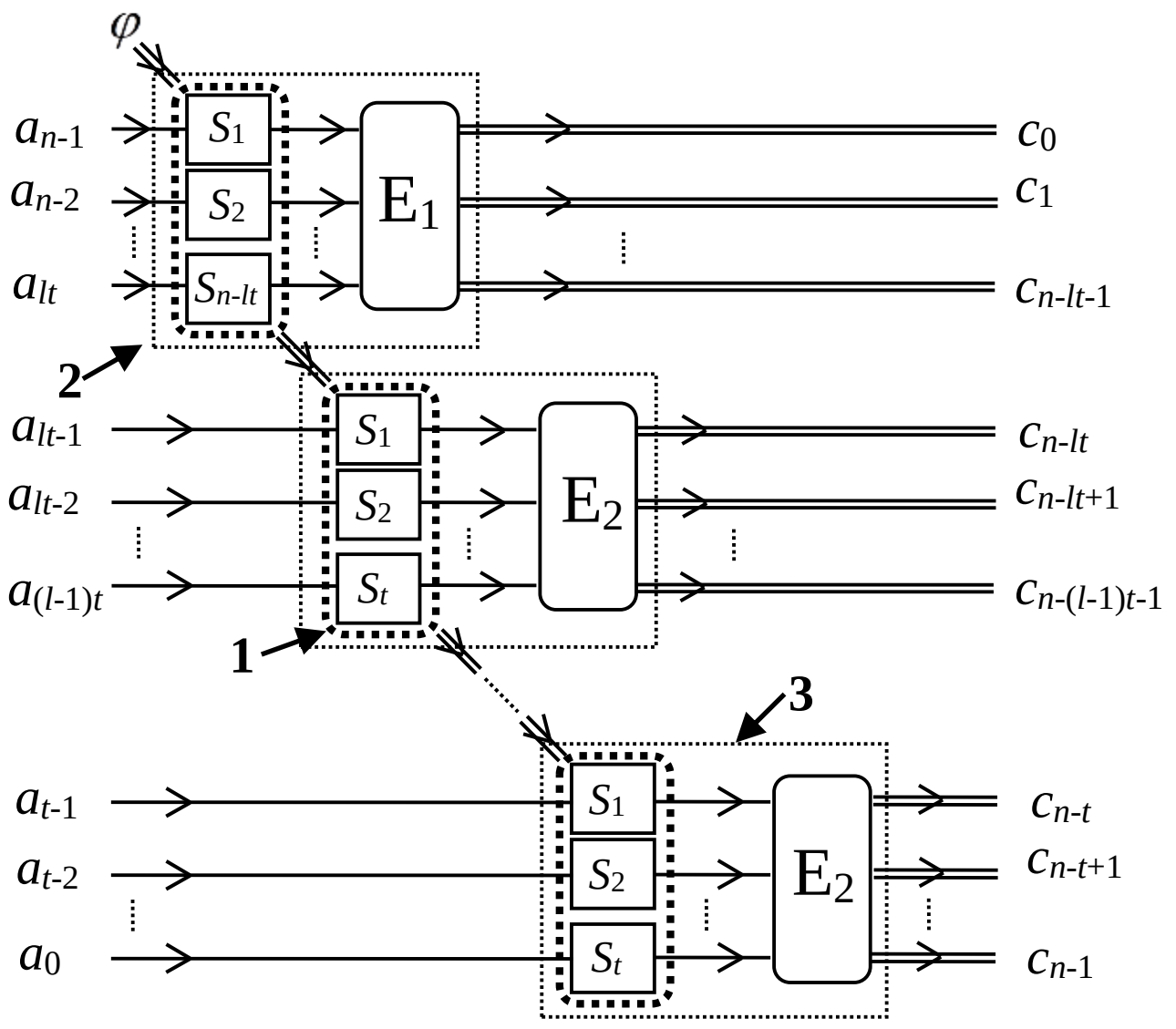

Fig.1 the circuit for $t$ –bit semiclassical QFT over $Z_{2^n}$

The explanation of Fig.1 is shown as follows:

(1) Each virtual box with bold line denotes VB1 while the one with fine line which contains $E_1$ denotes VB2 and the one with fine line which contains $E_2$ denotes VB3, where $E_1$ is the QFT over $Z_{2^{n-lt}}$ and $E_2$ is the QFT over $Z_{2^t}$. The circuit for $E_1$ and $E_2$ can refer to Ref.[11];

(2) The outputs of VB2 have $n-lt+1$ values namely $c_{n-lt-1},c_{n-lt-2},\cdots,c_0$, $\varphi'=\frac{c_0}{2^{n-lt+1}}+\cdots+\frac{c_{n-lt-1}}{2^2}$; If VB3 enters as $a'_{t-1},a'_{t-2},\cdots,a'_0$, it leaves as $t+1$ values namely $c'_{t-1},c'_{t-2},\cdots,c'_0$, $\varphi'=\frac{\varphi}{2^t}+\frac{c'_0}{2^{t+1}}+\cdots+\frac{c'_{t-1}}{2^2}$; $S_k=\begin{bmatrix}1 & 0\\ 0 & e^{2\pi i\varphi/2^{k-1}}\end{bmatrix}$ for $k=1,2,\cdots,t$, where $\varphi'$ is sent to the next VB1, $\varphi$ is the input of VB1 and the very first VB1 uses $\varphi=0$;

(3) The single line represents quantum signal and double line represents classical signal;

(4) $c=\sum_{j=0}^{n-1}2^j c_j$ is the final measurement.

We now analyze the required resource of the circuit for *t*-bit semiclassical QFT over $Z_{2^n}$.

Like single-bit semiclassical QFT, $t$ -qubit quantum register can be applied cyclically to prepare

$\left|a_{n-1},\cdots,a_{lt}\right\rangle,\left|a_{lt-1},\cdots,a_{(l-1)t}\right\rangle,\cdots,\left|a_{t-1},\cdots,a_0\right\rangle$. Thus, the *t*-bit semiclassical QFT over $Z_{2^n}$ requires $t$ -qubit quantum register. Since the QFT over $Z_{2^n}$ needs $n^2/2$ two-qubit gates and $n$ single-qubit gates[15], our circuit in Fig.1 needs $t^2/2$ two-qubit gates and $t$ single-qubit gates to realize QFT over $Z_{2^t}$. It also needs $t$ single-qubit gates to realize $R_j$ ( $j=1,2,\cdots,t$ ). Therefore, the number of the two-qubit gates and single-qubit gates for the *t*-bit semiclassical QFT are respectively $1/l^2$ and $2/l$ times as many as those of the QFT over $Z_{2^n}$. It is obvious that the *t*-bit semiclassical QFT requires the preparation of superposition $\frac{1}{2^{(n-l't)/2}}\sum_{c=0}^{2^{n-l't}-1}|c\rangle$ once and that of $\frac{1}{2^{t/2}}\sum_{c=0}^{2^t-1}|c\rangle$ $l-1$ times while the single-bit semiclassical QFT requires the preparation of superposition $(|0\rangle+|1\rangle)/\sqrt{2}$ $n$ times. Therefore, the running speed of *t*-bit semiclassical QFT is $t$ times that of single-bit semiclassical QFT.

## 4 Realization of *t*-bit semiclassical QFT over $Z_{2^n}$

In 2016, IBM opened the first 5 qubits quantum cloud computer[16]. We will study how to implement the QFT on IBM's quantum cloud computer, which provides limited quantum gates as follows.

$$\mathrm{U}_1(\lambda)=\begin{bmatrix}1 & 0\\ 0 & e^{\lambda i}\end{bmatrix},\ \mathrm{U}_2(\lambda,\varphi)=\frac{1}{\sqrt{2}}\begin{bmatrix}1 & \text{-}e^{\lambda i}\\ e^{\varphi i} & e^{(\lambda+\varphi)i}\end{bmatrix},\ \mathrm{U}_3(\theta,\lambda,\varphi)=\begin{bmatrix}\cos\frac{\theta}{2} & \text{-}e^{\lambda i}\sin\frac{\theta}{2}\\ e^{\varphi i}\sin\frac{\theta}{2} & e^{(\lambda+\varphi)i}\cos\frac{\theta}{2}\end{bmatrix},$$

$$\oplus=\begin{bmatrix}1&0&0&0\\0&1&0&0\\0&0&0&1\\0&0&1&0\end{bmatrix},\ \mathrm{H}=\frac{1}{\sqrt{2}}\begin{bmatrix}1&1\\1&-1\end{bmatrix},\ \mathrm{S}=\begin{bmatrix}1&0\\0&i\end{bmatrix}\ \text{and}\ \mathrm{T}=\begin{bmatrix}1&0\\0&(1+i)/\sqrt{2}\end{bmatrix}$$

The controlled- $S_k$ gate is one of the most important gates used in Fig.1, which can be realized by combination of multiple controlled- $R_j$ gates, however, $\oplus$ is the only two-qubit gate on IBM's quantum cloud computer. Thus we design a circuit for controlled- $R_k$ gate, according to the circuit for arbitrary controlled- $U$ gate by single-qubit gate and $\oplus$ gate in Ref.[15], which is shown as Fig.2.

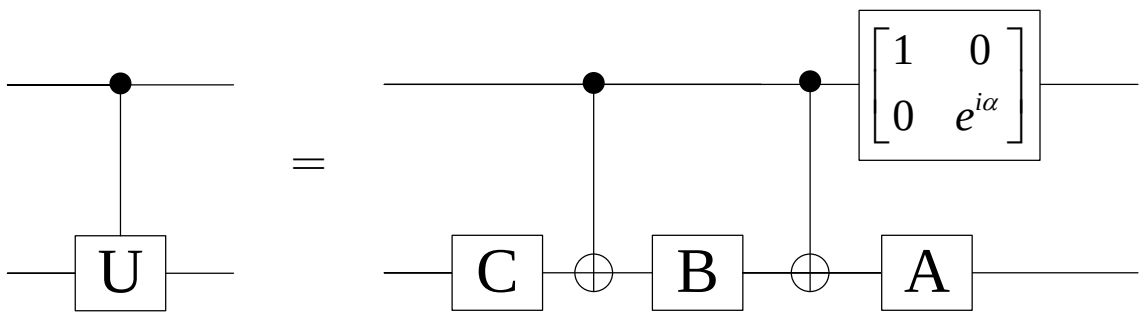

Fig.2 the circuit for controlled- $U$ gate

Note that, in Fig.2, $\alpha$ , A, B and C satisfy $U=e^{i\alpha}AXBXC$ and $ABC=I$, where A, B, and C are all single-qubit gate, $X=\begin{bmatrix}0&1\\1&0\end{bmatrix}$ and $I=\begin{bmatrix}1&0\\0&1\end{bmatrix}$.

Taking controlled- $R_4$ gate as an example, which can be realized by $\mathrm{U}_1(\pi/32)$ , $\mathrm{U}_1(-\pi/16)$ , $\mathrm{U}_1(\pi/32)$ and $\alpha=\pi/16$ (based on Matlab software). And the circuit is shown as Fig.3.

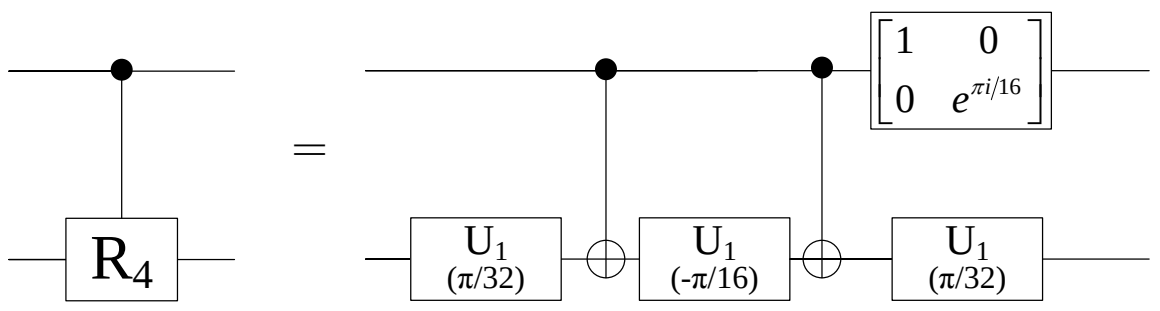

Fig.3 the circuit for the controlled- $R_4$ gate on IBM's quantum cloud computer

When the input state is $|00\rangle$, the running result of Fig.3 is shown in Fig.4. To quantitatively evaluate the performance, we use the squared statistical overlap [17] of experimental data with the ideal values, which is defined as $\gamma = \left(\sum_{y=0}^{3} \sqrt{m_y e_y}\right)^2$, where $m_y$ and $e_y$ are the measured and expected output-state probabilities of the state $|y\rangle$ respectively. From the data in Fig.4, we find $\gamma = 0.956$, indicating a near perfect experimental accuracy. Thus, the circuit in Fig.3 can realize the controlled- $R_4$ gate.

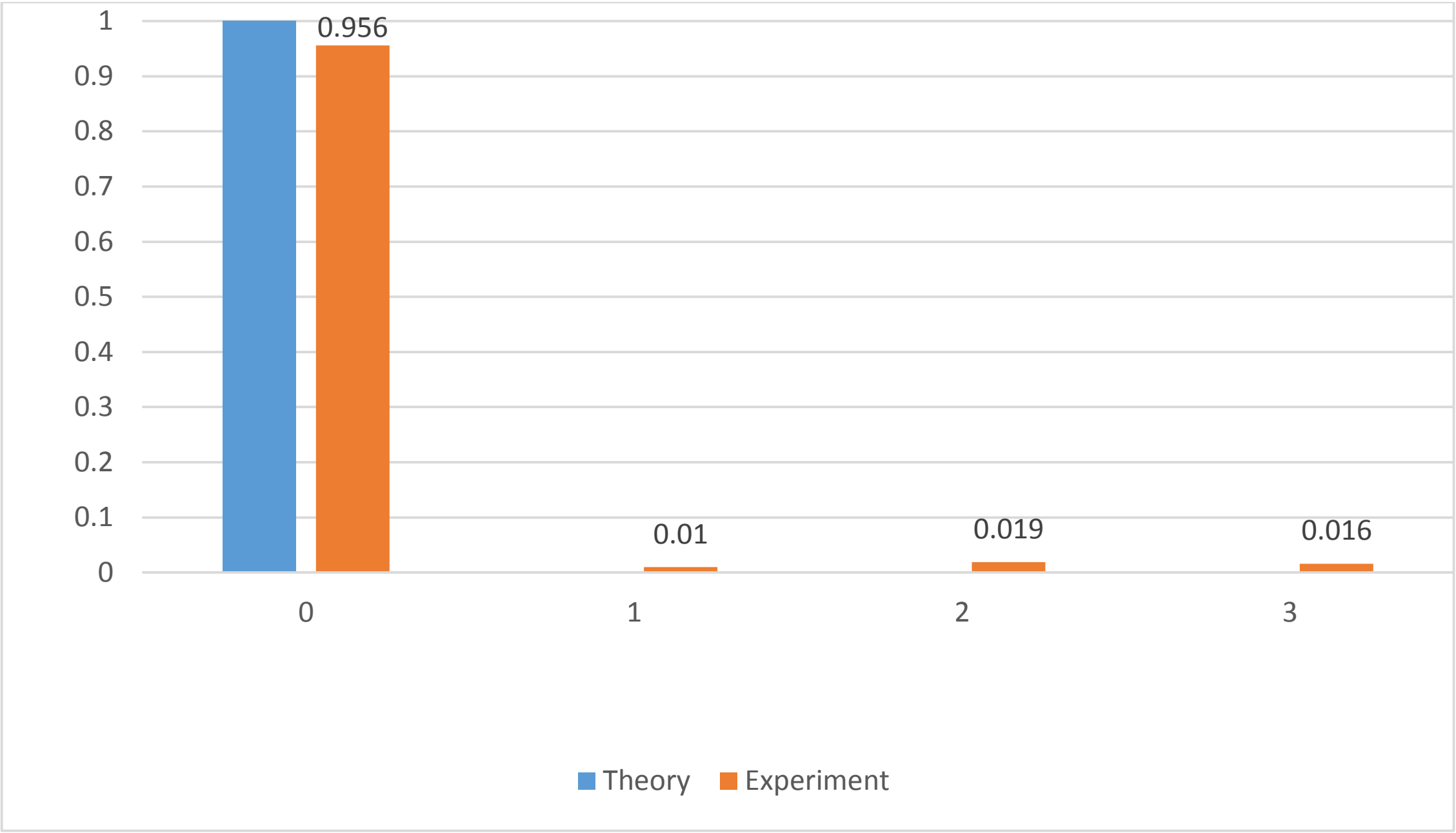

Fig.4 the running result of the controlled- $R_4$ gate

Based on the circuit in Fig.3, we design the method to realize it on IBM's quantum cloud computer, which verifies the feasibility (taking 2-bit semiclassical QFT over $Z_{2^3}$ as an example).

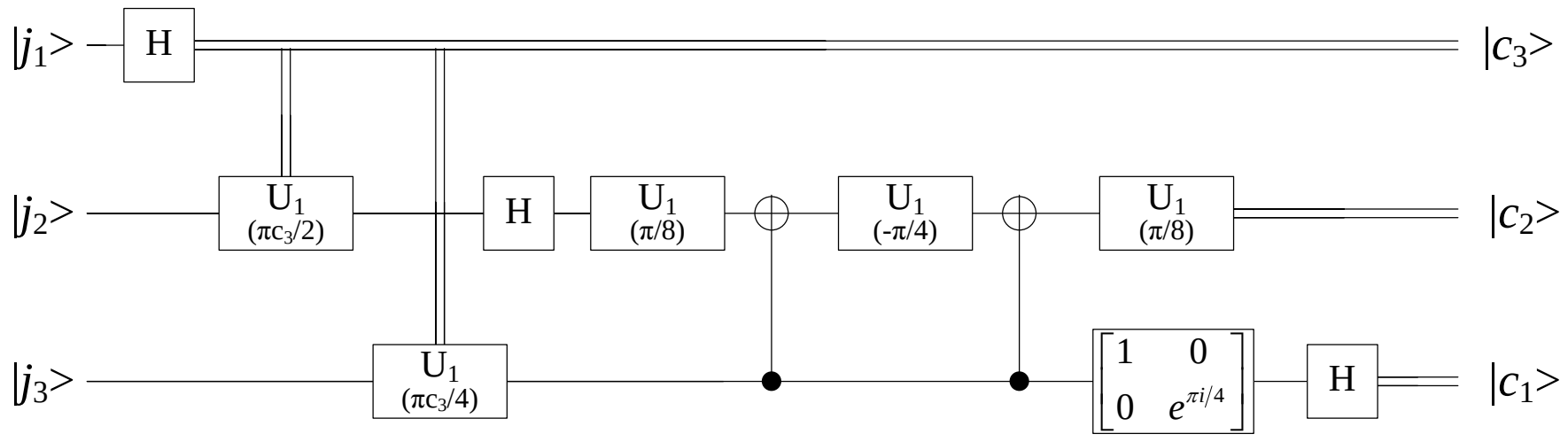

Fig.5 the circuit for 2-bit semiclassical QFT over $Z_{2^3}$ on IBM's quantum cloud computer

When the input state is $|j_1 j_2 j_3\rangle = |000\rangle$, the running result of the above circuit is shown in Fig.6. And the squared statistical overlap of experimental data with the ideal values is $\gamma' = 0.9952$, indicating a near perfect experimental accuracy. Thus, the circuit in Fig.5 can realize 2-bit semiclassical QFT over $Z_{2^3}$, i.e. $t$ -bit semiclassical QFT over $Z_{2^n}$ can be applied to realize the QFT over $Z_{2^n}$.

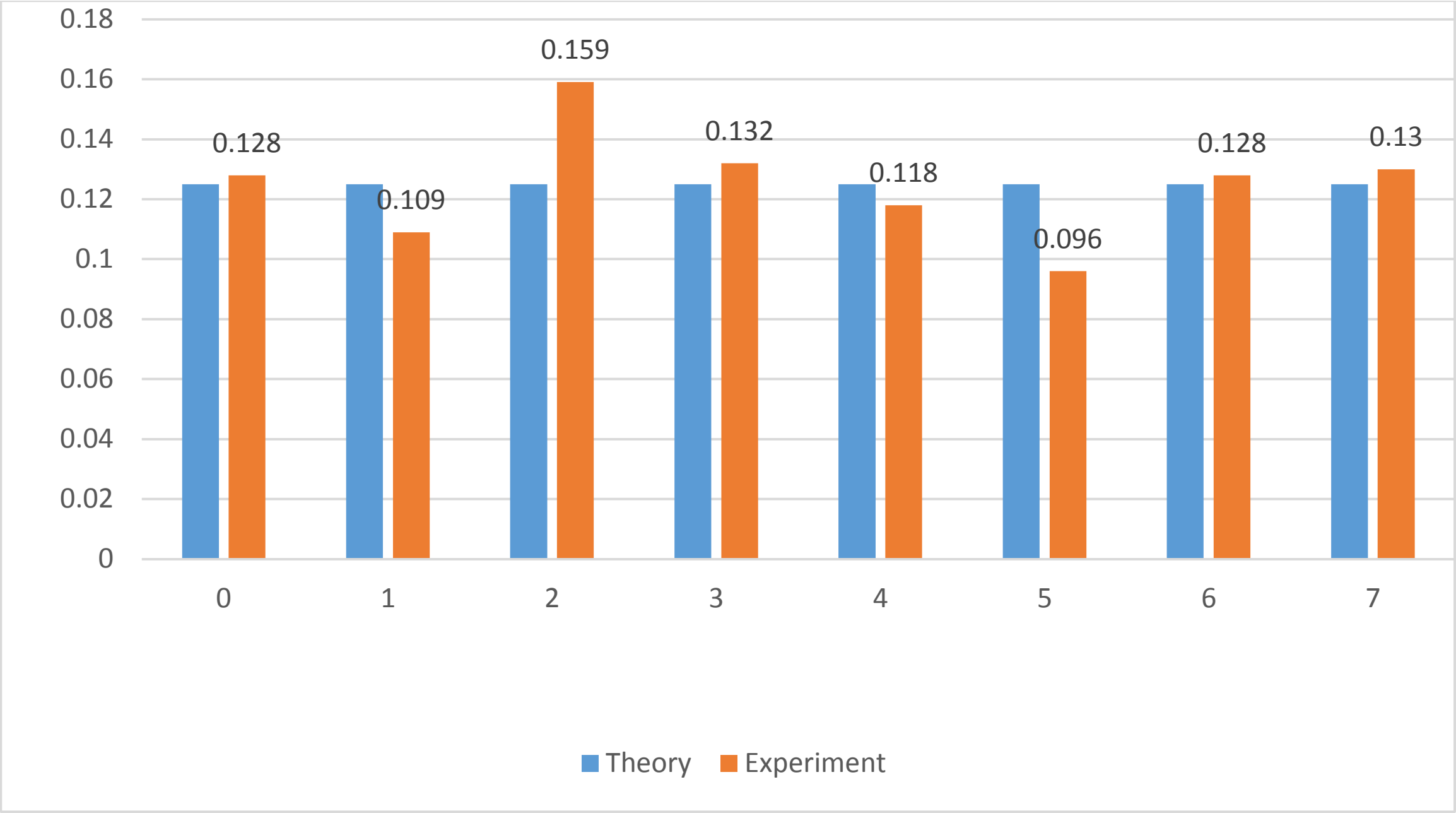

Fig.6 the running result of 2-bit semiclassical QFT over $Z_{2^3}$

And $t$-bit semiclassical QFT over $Z_{2^n}$ is theoretically superior to the QFT over $Z_{2^n}$. For the sake of comparison, we will realize the QFT over $Z_{2^n}$. And the circuit for the QFT over $Z_{2^3}$ can be designed as Fig.7, which can be implemented on IBM's quantum cloud computer.

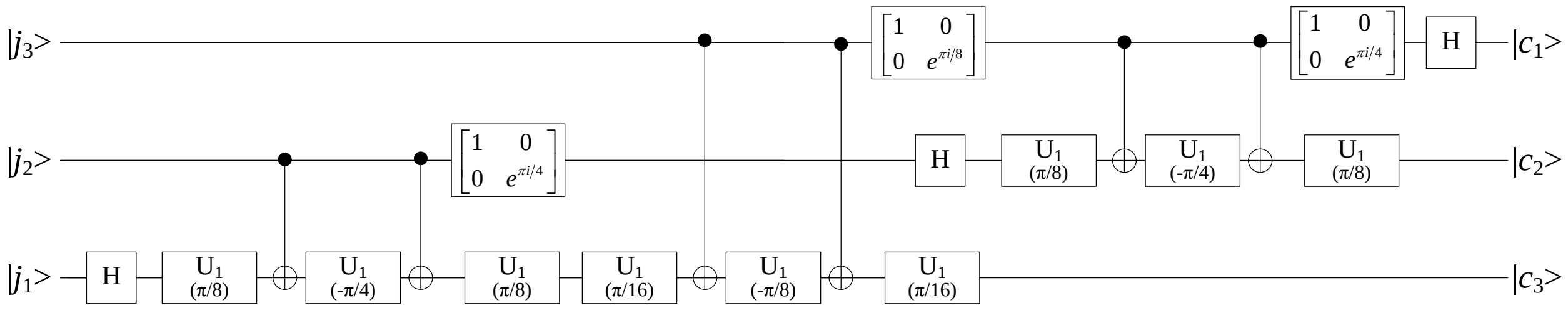

Fig.7 the circuit for the QFT over $Z_{2^3}$ on IBM's quantum cloud computer

When the input state is $|j_1 j_2 j_3\rangle = |000\rangle$, the running result of the above circuit is shown in Fig.8. And the squared statistical overlap of experimental data with the ideal values is $\gamma'' = 0.9355$ which is lower than $\gamma'$, indicating a near perfect experimental accuracy. And Fig.5 needs 2 $\oplus$ gates, which is less than Fig.7, i.e. the fidelity of Fig.5 is higher than Fig.7 (the less the two-qubit gate is needed, the higher the fidelity is). Thus the fidelity and squared statistical overlap of Fig.5 are both higher than Fig.7, i.e. the circuit in Fig.1 is superior to standard QFT.

The QFT over $Z_{2^3}$ in Fig.7 needs 3-qubit quantum register, while 2-bit semiclassical QFT over $Z_{2^3}$ in Fig.5 needs only 2-qubit quantum register. Furthermore, from Fig.1 and Fig.5, we can see that it takes only 2-qubit quantum register to realize the QFT over $Z_{2^n}$ based on the semiclassical method. In other words, the method shows a good scalability. That is, $t$-bit semiclassical QFT over $Z_{2^n}$ provides a solution for implementing a large-scale QFT. And the QFT over $Z_{2^n}$ can be realized on IBM's quantum cloud computer by $t$-bit semiclassical QFT over $Z_{2^n}$, where $t \le 5$.

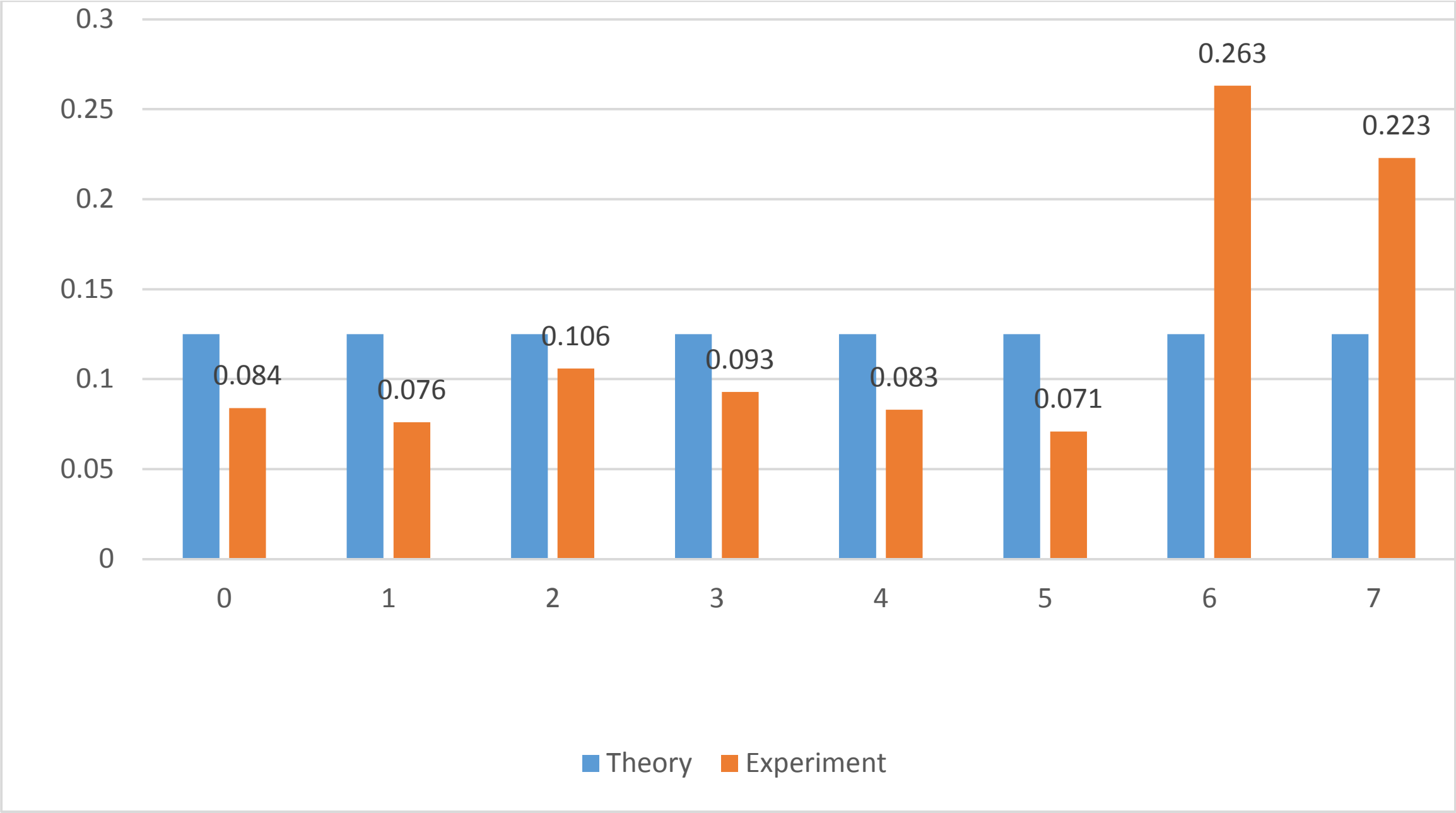

Fig.8 the running result of the QFT over $Z_{2^3}$

## 5 Realization of Shor's quantum algorithm

In this section, based on *t*-bit semiclassical QFT, we will realize Shor's algorithm, which can further demonstrate the feasibility of the method. And the application process of Shor's algorithm will be accelerated under the condition of limited computation resources.

Shor's algorithm was presented in 1994, which can break through the RSA public-key crypto. The strategy for quantum factoring a composite number $N = pq$ , with both *p* and *q* being odd primes, is as follows[1].

First, we choose $x$ which is coprime to *N*.

Next, two registers of qubits are used: the first register with $n = 2\lceil \log_2 N \rceil$ qubits and the second with $l = \lceil \log_2 N \rceil$ qubits, which are initialized in the quantum state $|0\rangle^{\otimes n} |0\rangle^{\otimes l}$ . The first register becomes $\frac{1}{2^{n/2}} \sum_{j=0}^{2^n - 1} | j\rangle |0\rangle$ after applying QFT over $Z_{2^n}$ .

Next, computing modular exponential function $x^j \bmod N$ in the second register, it will become $\frac{1}{2^{n/2}} \sum_{j=0}^{2^n - 1} | j\rangle \left| x^j \bmod N \right\rangle$.

Then, applying QFT over $Z_{2^n}$ on the first register, we can obtain $\frac{1}{2^n} \sum_{j=0}^{2^n - 1} \sum_{c=0}^{2^n - 1} e^{2\pi ijc/2^n} |c\rangle \left| x^j \bmod N \right\rangle$.

Finally, we observe the machine. And we will obtain solely $c$ in the first register.

Shor proved that period $r$ can be obtained by using a continued fraction expansion of $c/2^n$ , when $\left| \frac{c}{2^n} - \frac{d}{r} \right| \le \frac{1}{2^{n+1}}$ , with period $r$ being the smallest positive satisfying $x^r \bmod N = 1$ and $d = \left[ rc/2^n \right]$ . Based on this period *r*, at least one nontrivial factor of *N* can be given by the greatest common divisor (GCD) of $(x^{r/2} \pm 1) \bmod N$ with probability greater than $1/2$ [15].

It is apparent that Shor's factorization algorithm needs $3\lceil \log_2 N \rceil$-qubit quantum register, where the QFT needs $2\lceil \log_2 N \rceil$-qubit quantum register. For $N = 15$, the QFT over $Z_{2^8}$ in Shor's algorithm can't be realized on IBM's 5 qubits quantum cloud computer. Luckily, the QFT can also be realized by the *t*-bit semiclassical QFT, then Shor's factorization algorithm can be realized. The circuit is designed as Fig.9, where $x = 11$ and $t = 2$.

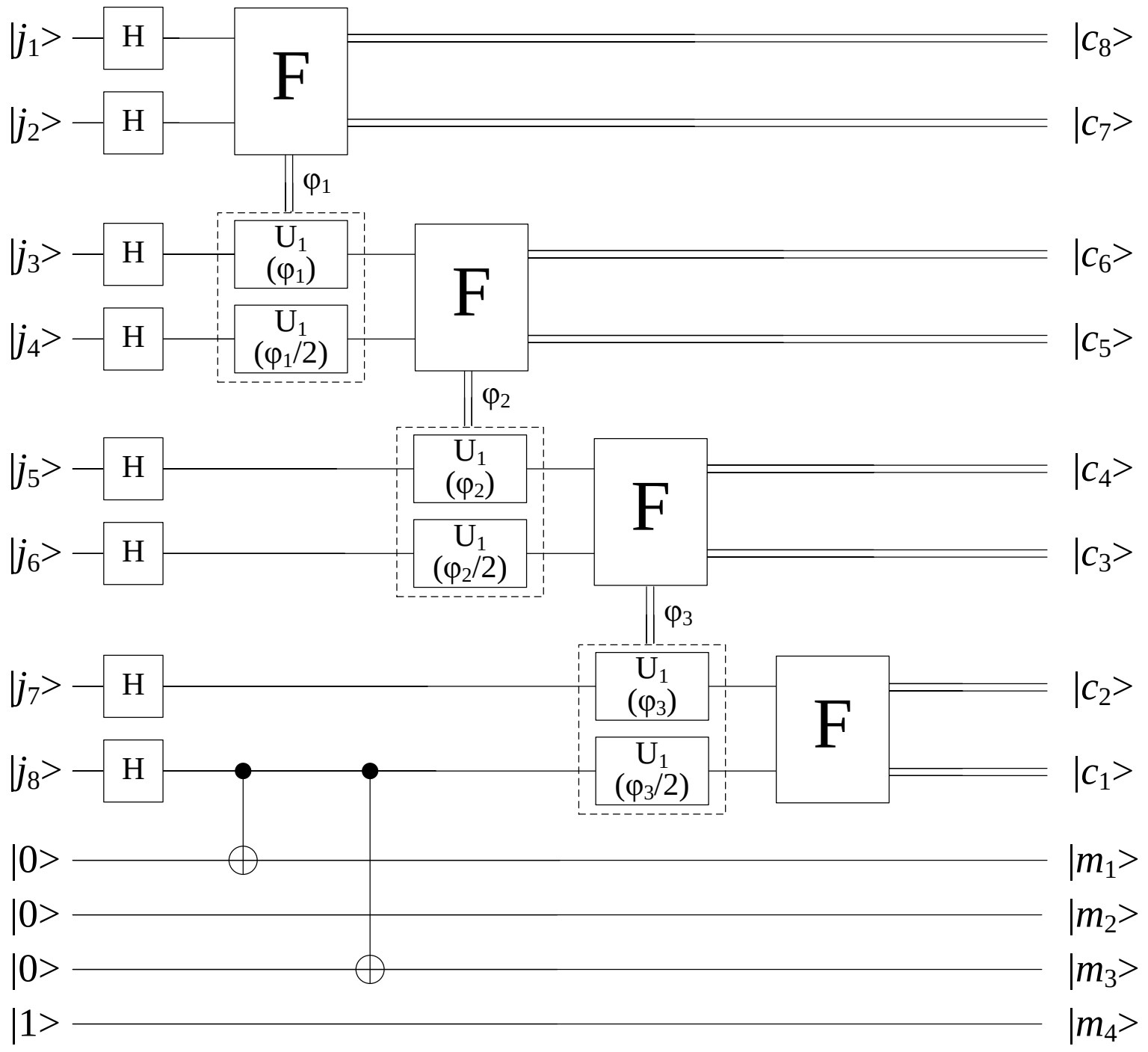

Fig.9 the circuit for Shor's quantum algorithm based on the *t*-bit semiclassical QFT

The explanation of Fig.9 is shown as follows:

(1) $\varphi_1 = (\frac{c_8}{4} + \frac{c_7}{2})\pi$, $\varphi_2 = (\frac{c_8}{16} + \frac{c_7}{8} + \frac{c_6}{4} + \frac{c_5}{2})\pi$, $\varphi_3 = (\frac{c_8}{64} + \frac{c_7}{32} + \frac{c_6}{16} + \frac{c_5}{8} + \frac{c_4}{4} + \frac{c_3}{2})\pi$.

(2) The circuit of $11^a \bmod 15$ is quoted from Ref.[18], which only needs 2-qubit quantum register.

(3) Like the single-bit semiclassical QFT, 2-qubit quantum register can be applied cyclically to prepare $|j_1 j_2\rangle, \cdots, |j_7 j_8\rangle$. That is to say the circuit can be implemented in 4 steps. The first three steps respectively needs 2-qubit quantum register, the last step 4-qubit quantum register.

(4) The circuit for F is shown in Fig.10.

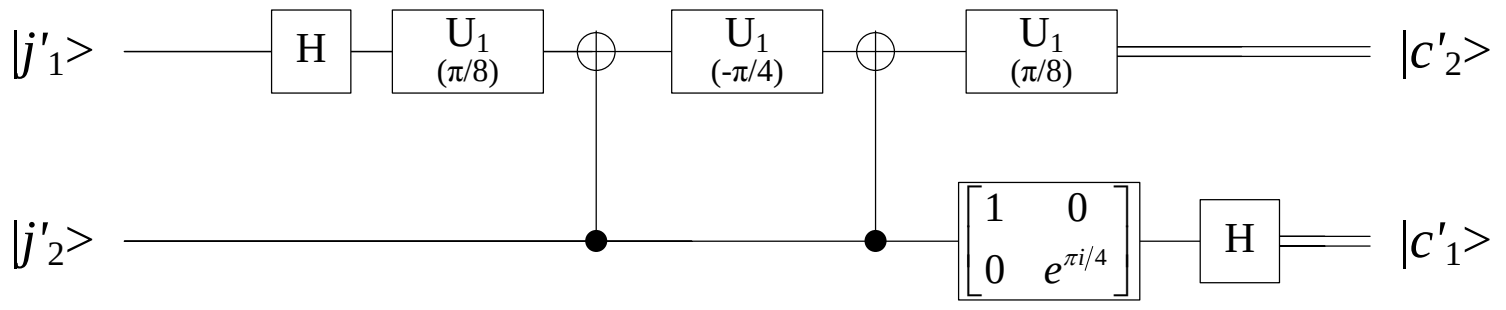

Fig.10 the circuit for F

And the experimental result is shown as Fig.11.

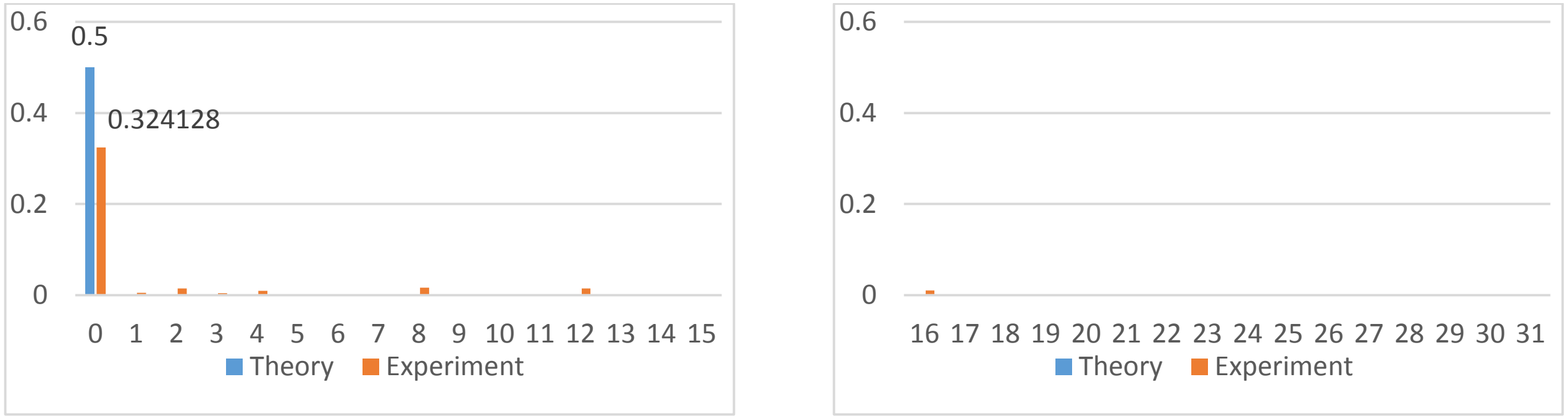

(1) (2)

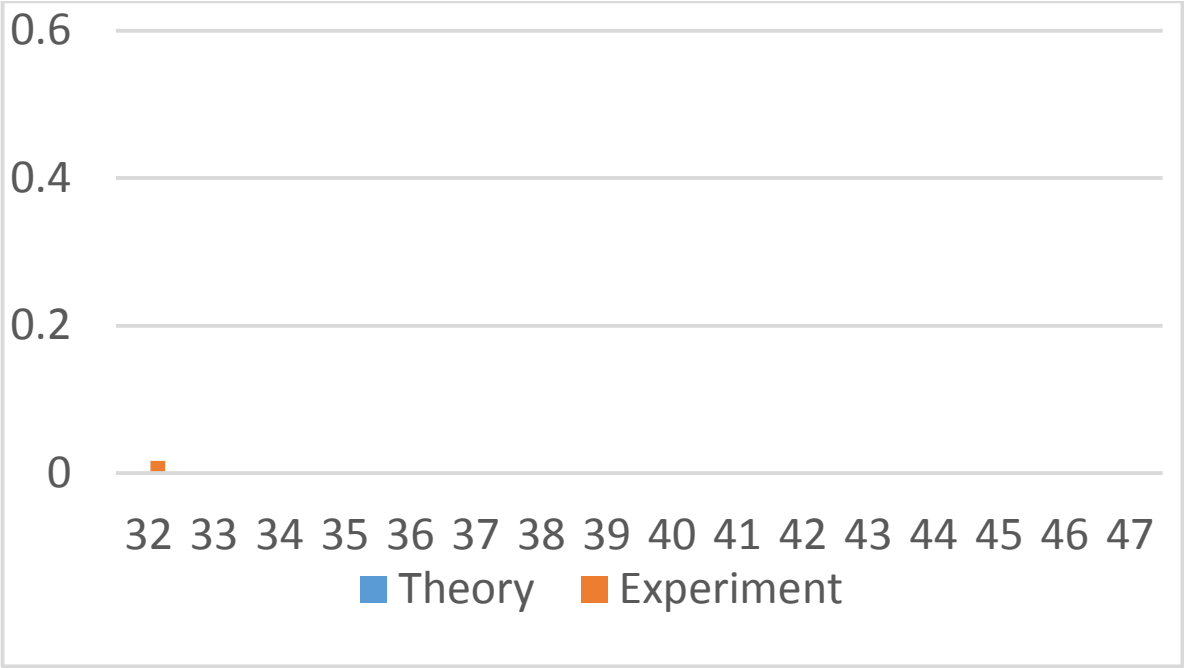

(3)

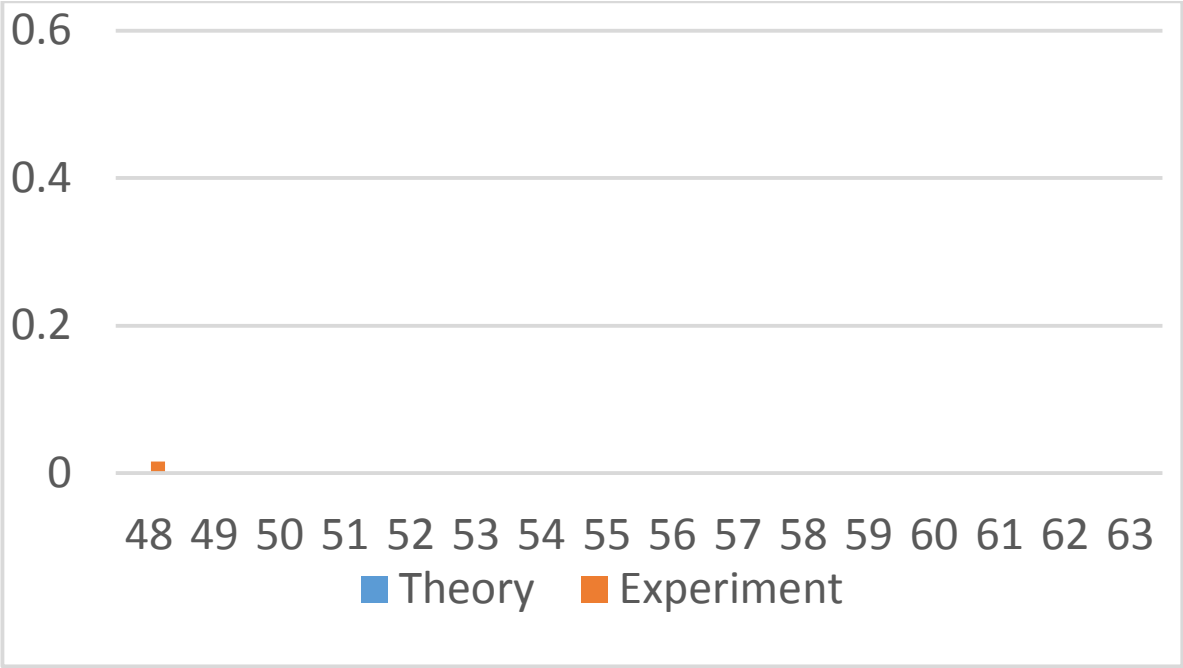

(4)

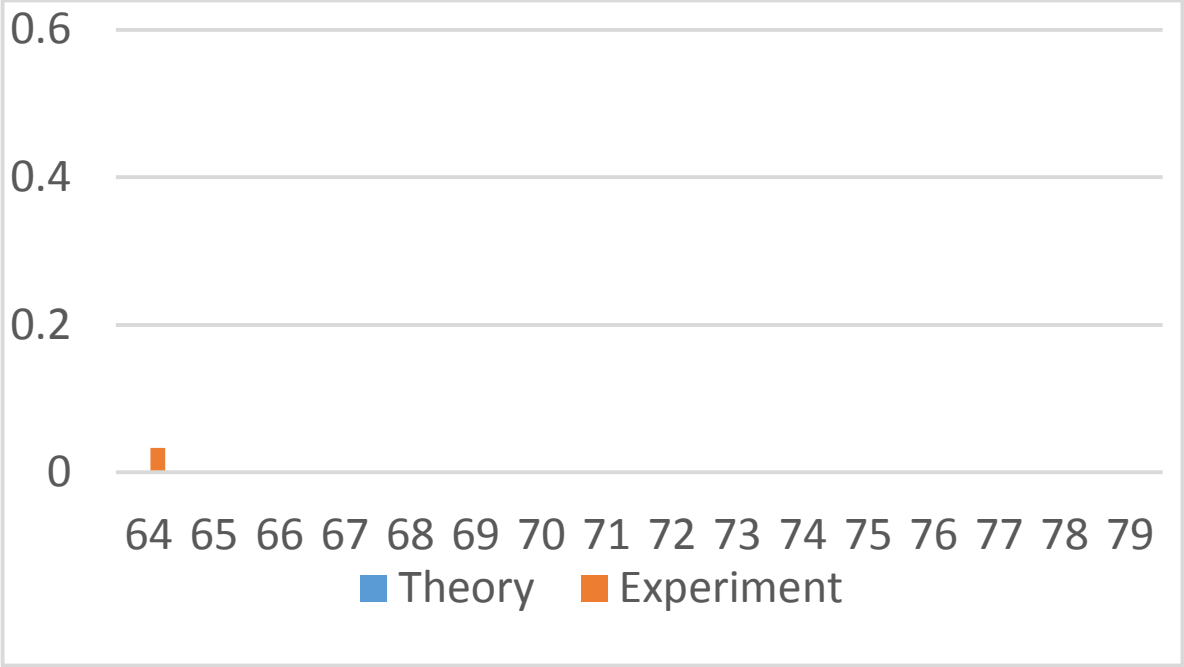

(5)

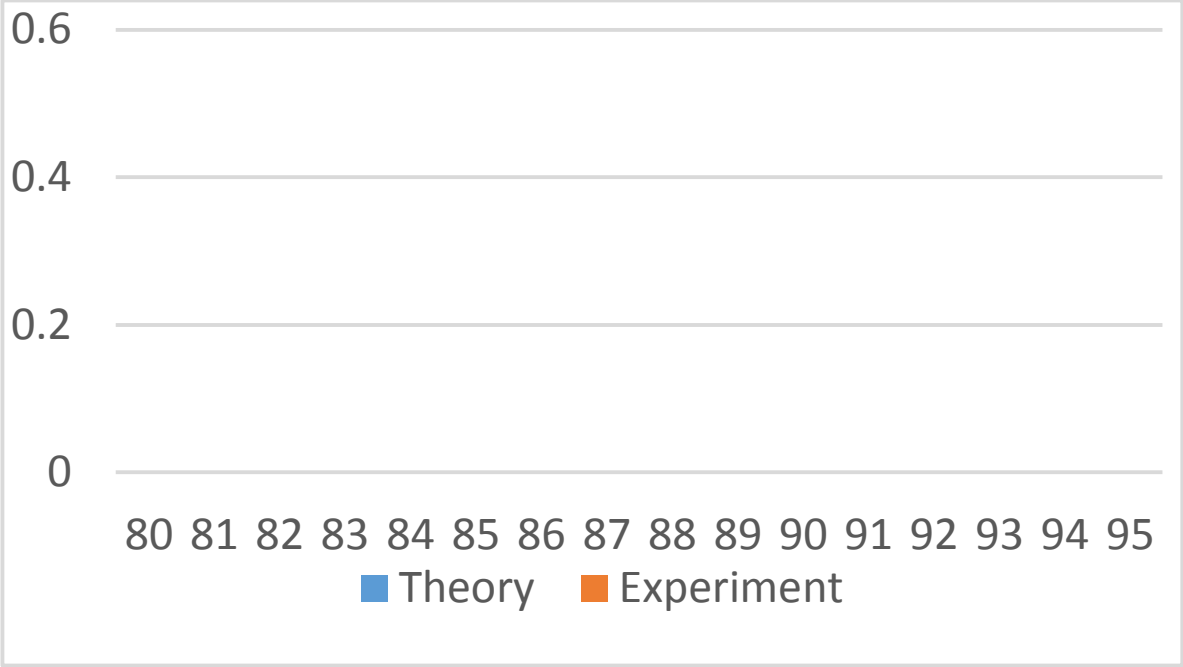

(6)

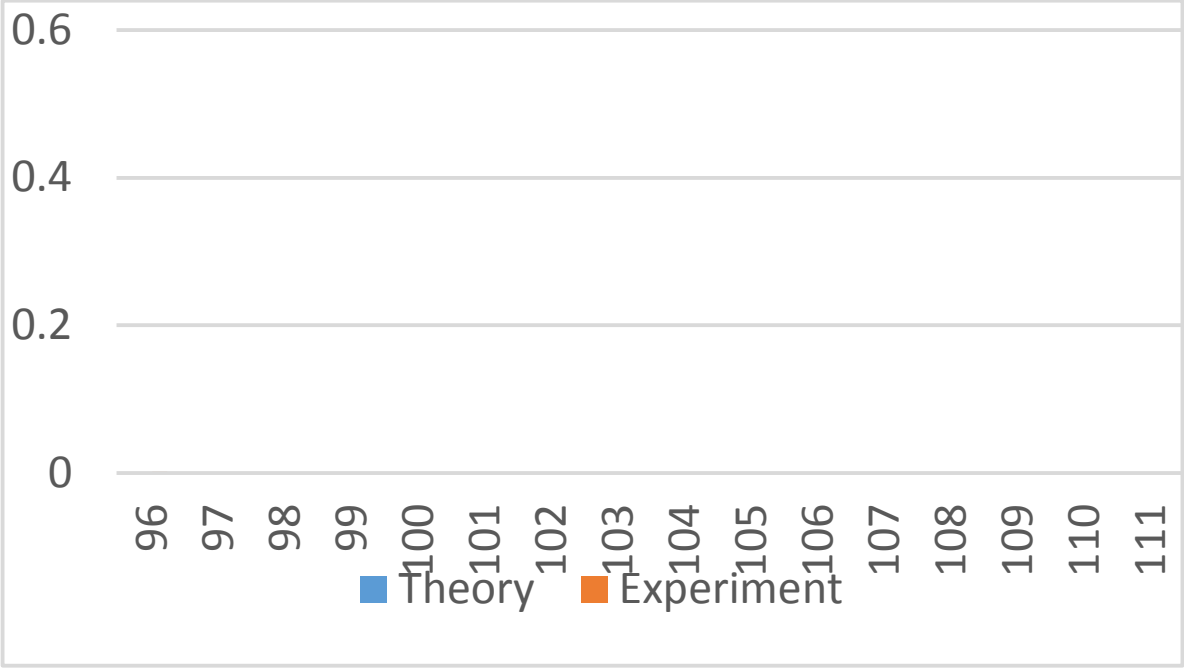

(7)

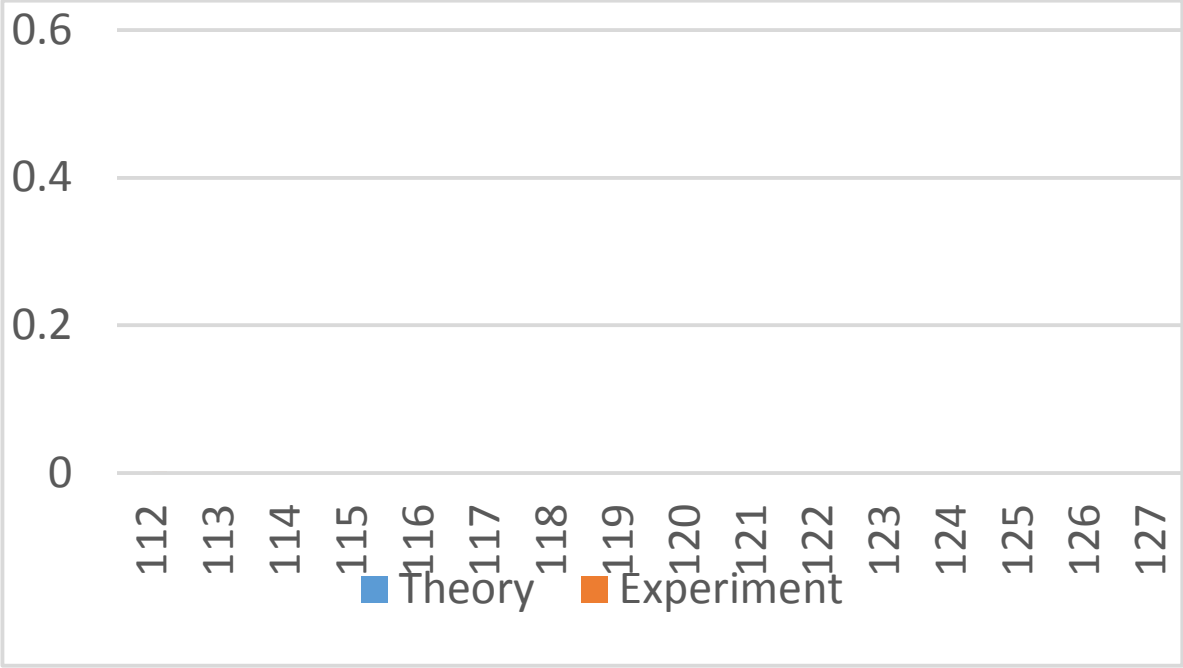

(8)

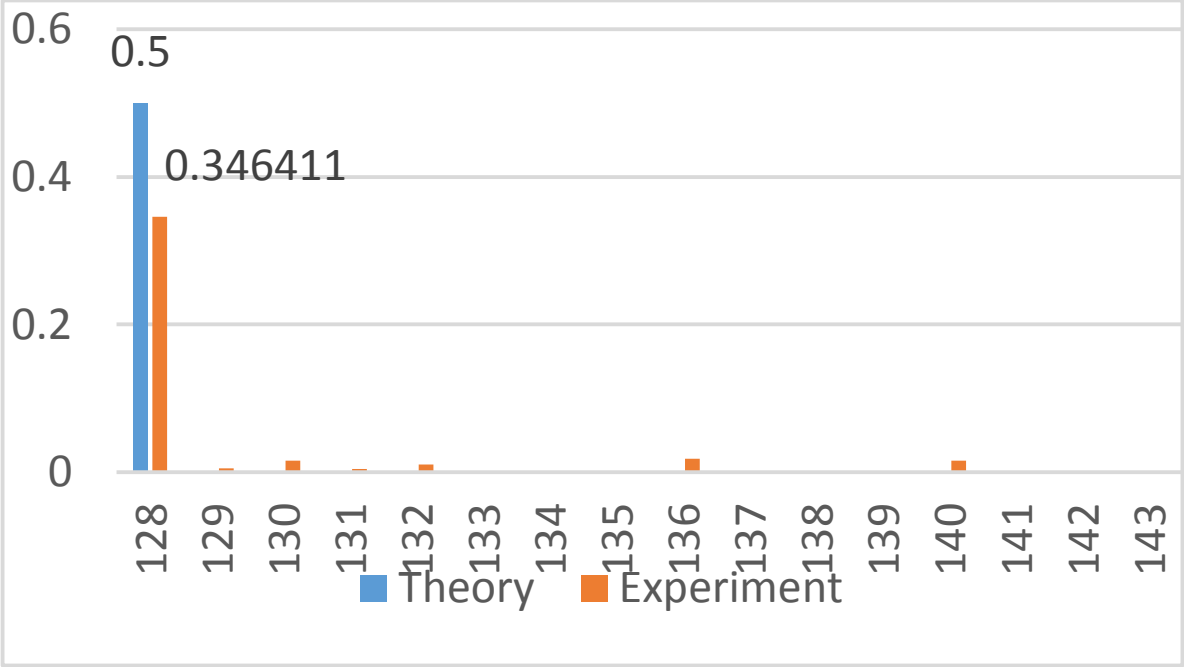

(9)

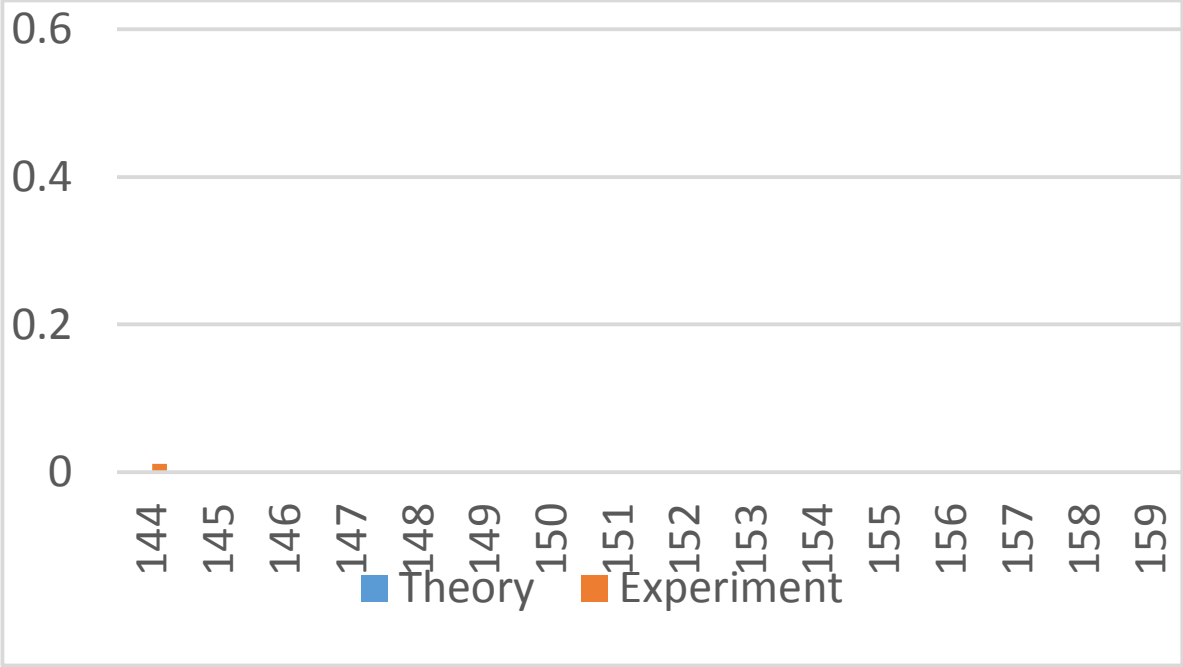

(10)

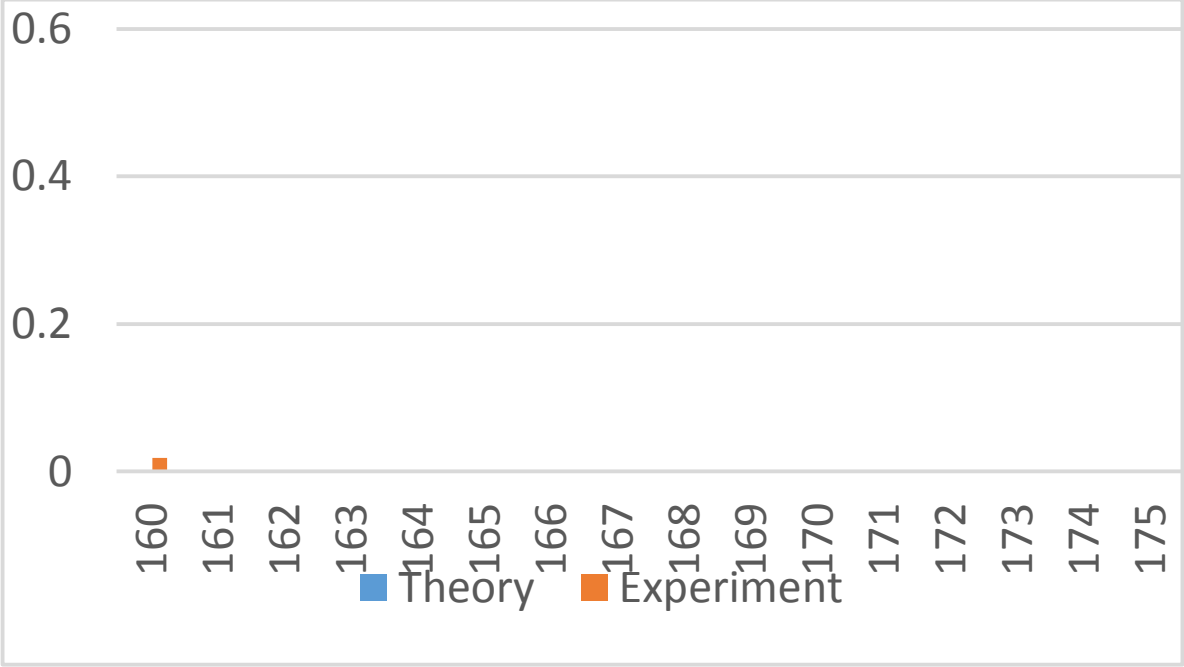

(11)

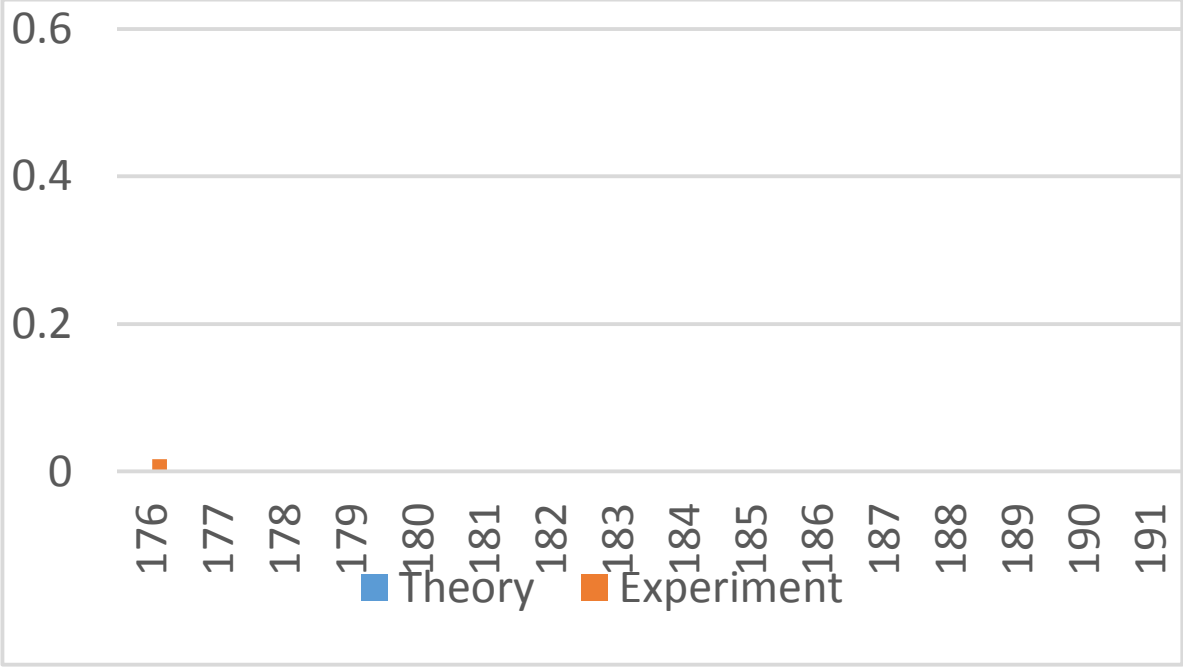

(12)

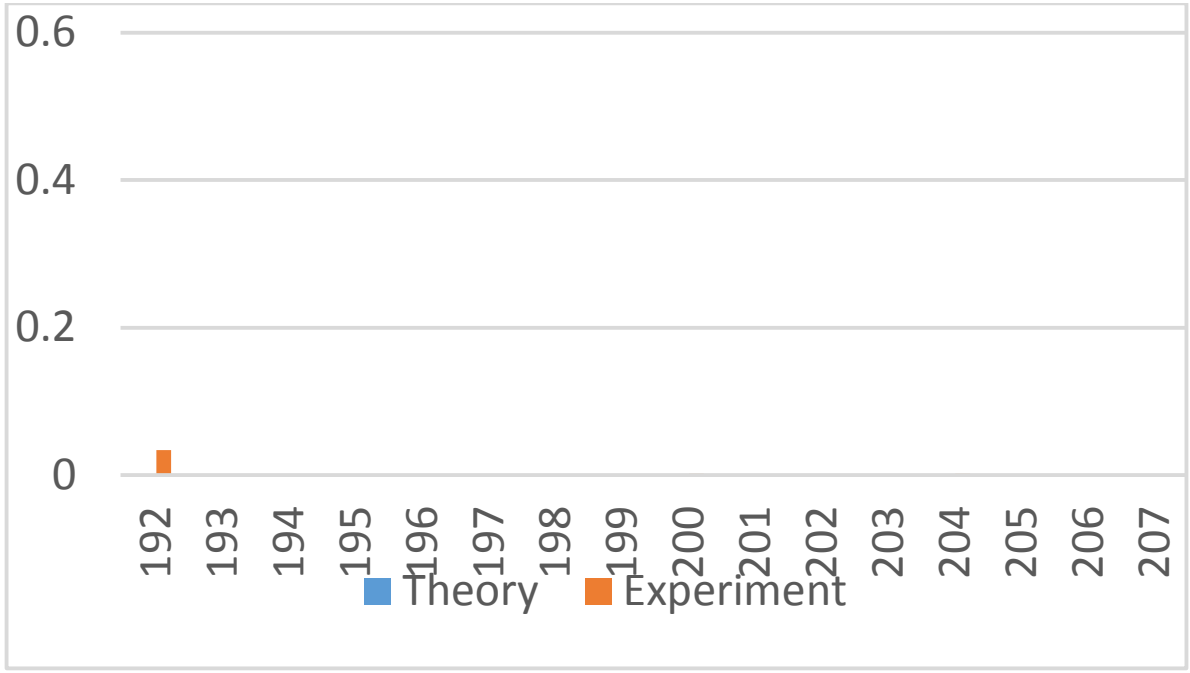

(13)

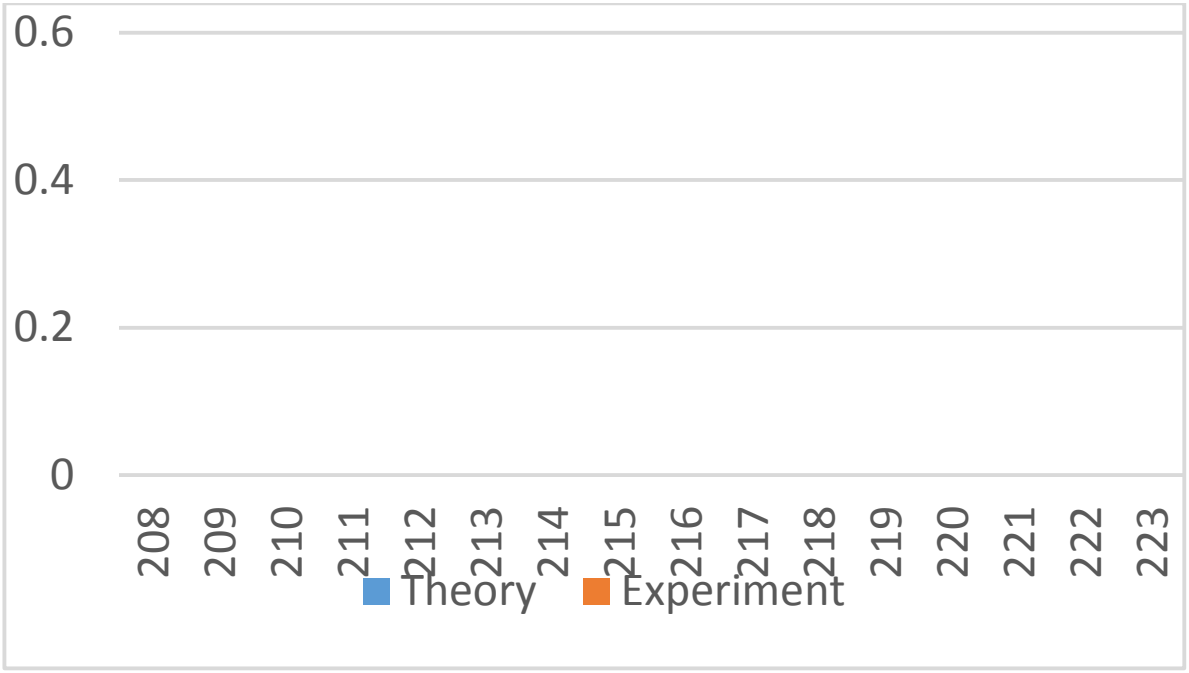

(14)

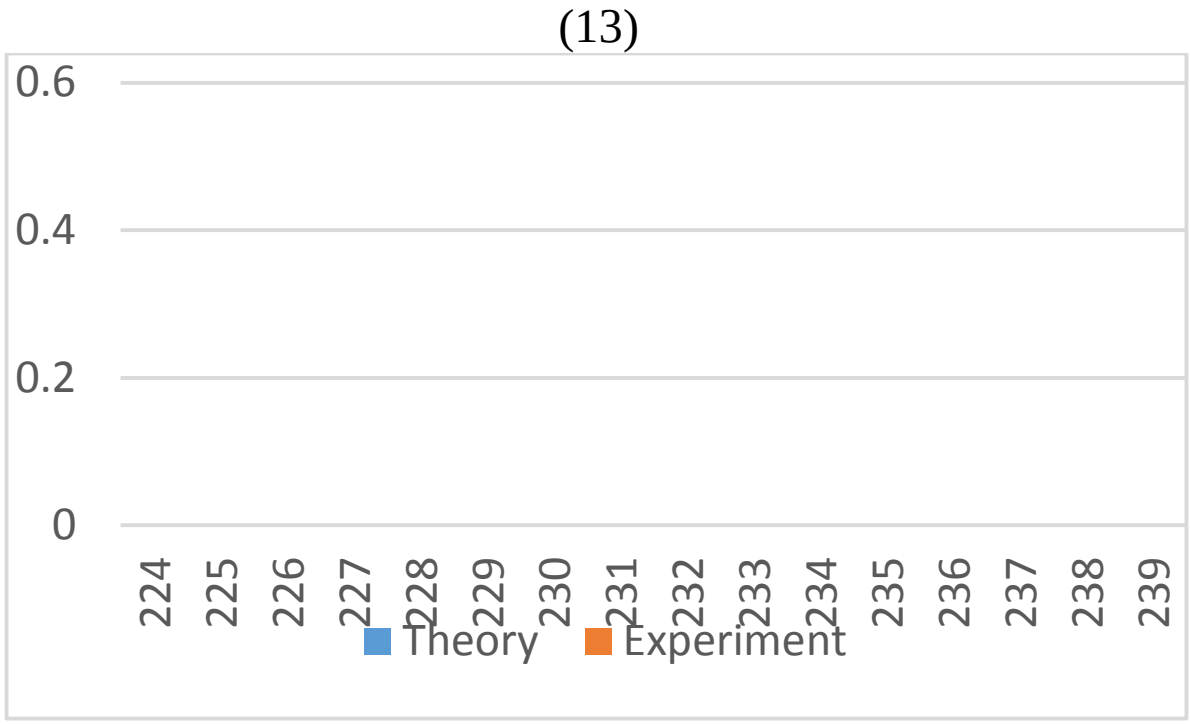

(15)

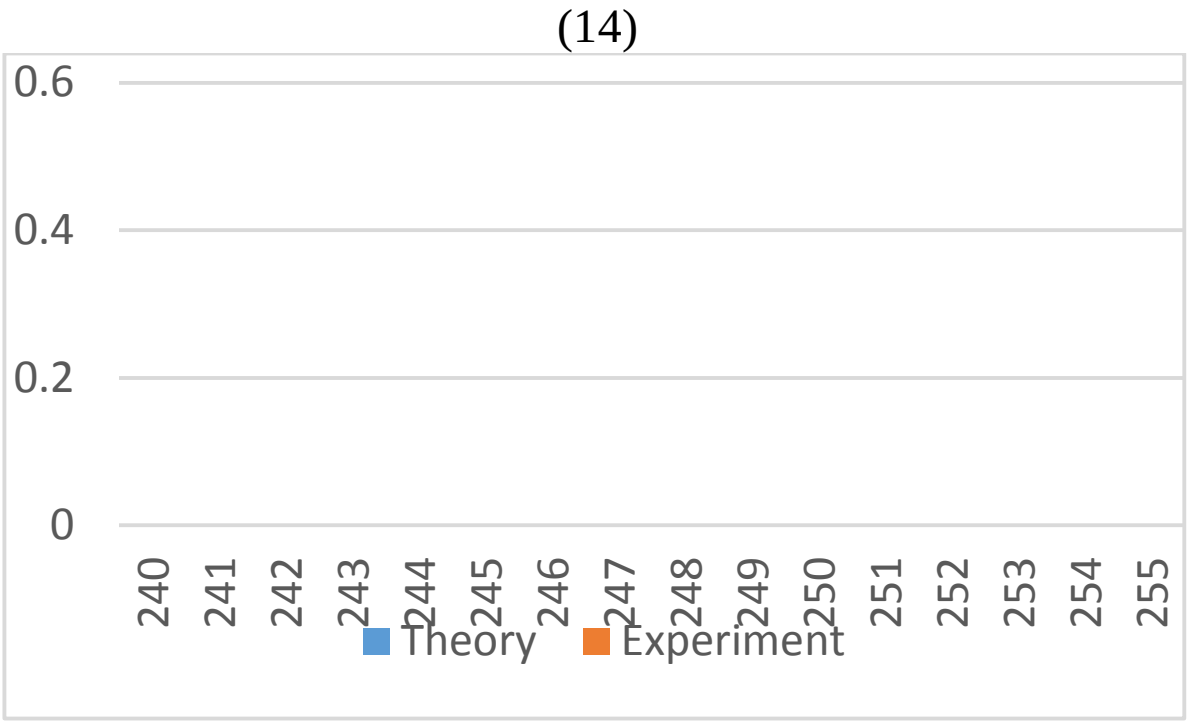

(16)

Fig.11 the running result of Shor's quantum algorithm based on the *t*-bit semiclassical QFT. And on the horizontal axis is the result $c=\sum_{k=1}^{8}2^{8-k}c_k$ , on the vertical the measuring probability.

According to Fig.11, the experiments yields 128 with probability 0.346411, with $r=2$ being determined by the continued fraction expansion of $128/256$, the $\text{GCD}(11^{2/2}\pm1,15)=3,5$, giving a successful factorization. While, the experiments also yields 0 with probability 0.324128 and other values with probabilities very close to 0, corresponding to a failure.

In 2017, IBM announced that they built the operational prototype 50-qubit processor[19], which has more ability to implement Shor's algorithm. Then, based on the *t*-bit semiclassical QFT, the $N$ with the maximum digit 49 can be factorized for $t=1$, however, the maximum digit of $N$ is only 17 based on the standard QFT. Thus, the *t*-bit semiclassical QFT has significant advantage over the standard QFT, which will accelerate the application process of Shor's algorithm.

## 6 Conclusions

In this paper, we defined the *t*-bit semiclassical QFT, which can balance the tradeoff of the quantum resources required and the speed of implementation of the QFT. Then, we realize the 2-bit semiclassical QFT over $Z_{2^3}$ on IBM's quantum cloud computer. And the method can, in principle, be scaled to realize larger-scale QFT. Thus, this scheme provides a solution for implementing a large-scale QFT under the condition of limited computational resources. And based on the method, Shor's algorithm can be successfully implemented on IBM's quantum cloud computer.

## Acknowledge

The authors acknowledge the use of IBM's Quantum Experience for this work. This project is supported by the National Basic Research Program of China (Grant No.2013CB338002), National Natural Science Foundation of China (Grants No.61502526).